\documentclass[apj]{emulateapj}
\usepackage{natbib}     
\usepackage{fancyref}

\begin{document}

\title{Simultaneous Modeling of the Stellar and Dust Emission in Distant Galaxies: \\ Implications for Star Formation Rate Measurements}

\author{Dyas Utomo\altaffilmark{1}, Mariska Kriek\altaffilmark{1}, Ivo Labb{\'e}\altaffilmark{2}, Charlie Conroy\altaffilmark{3}, and Mattia Fumagalli\altaffilmark{2}}

\altaffiltext{1}{Astronomy Department, University of California, Berkeley, CA 94720, USA; dyas@berkeley.edu.}
\altaffiltext{2}{Leiden Observatory, Leiden University, P.O. Box 9513, NL-2300 RA Leiden, Netherlands.}
\altaffiltext{3}{Department of Astronomy \& Astrophysics, University of California, Santa Cruz, CA 95060, USA.}

\begin{abstract}
We have used near-ultraviolet (NUV) to mid-infrared (MIR) composite spectral energy distributions (SEDs) to simultaneously model the attenuated stellar and dust emission of $0.5 \lesssim z \lesssim 2.0$ galaxies. These composite SEDs were previously constructed from the photometric catalogs of the NEWFIRM Medium-Band Survey, by stacking the observed photometry of galaxies that have similar rest-frame NUV-to-NIR SEDs. In this work, we include a stacked MIPS 24$\micron$ measurement for each SED type to extend the SEDs to rest-frame MIR wavelengths. Consistent with previous studies, the observed MIR emission for most SED types is higher than expected from only the attenuated stellar emission. We fit the NUV-to-MIR composite SEDs by the Flexible Stellar Population Synthesis (SPS) models, which include both stellar and dust emission. We compare the best-fit star formation rates (SFRs) to the SFRs based on simple UV+IR estimators. Interestingly, the UV and IR luminosities overestimate SFRs -- compared to the model SFRs -- by more than $\sim$1\,dex for quiescent galaxies, while for the highest star-forming galaxies in our sample the two SFRs are broadly consistent.  The difference in specific SFRs also shows a gradually increasing trend with declining specific SFR, implying that quiescent galaxies have even lower specific SFRs than previously found. Contributions from evolved stellar populations to both the UV and the MIR SEDs most likely explain the discrepancy. Based on this work, we conclude that SFRs should be determined from modeling the attenuated stellar and dust emission simultaneously, instead of employing simple UV+IR-based SFR estimators.
\end{abstract}


\section{Introduction}

SEDs of galaxies contain a lot of information regarding their physical properties, such as the SFR, star formation history (SFH), metallicity, age of the stellar population, stellar mass, and the amount of dust \citep[e.g.,][]{conroy13}. These quantities can be extracted by fitting SEDs of galaxies with SPS models \citep[e.g.,][]{bruzual03,maraston05,dacunha08,conroy09}. SPS modeling is currently one of the most popular and powerful methods to study the SFHs and stellar mass build-up of galaxies over cosmic time \citep[e.g.,][]{labbe10b,wuyts11,brammer11,muzzin13}.

SPS models are generally combined with a dust model to account for dust attenuation of the stellar light. However, modeling just the attenuated stellar emission results in significant uncertainties on the derived dust content and consequently, other properties. Extending the stellar SEDs to IR wavelengths improves the constraints on the dust parameters, which in turn yields more accurate stellar population properties. The inclusion of IR data has primarily resulted in empirical SFR indicators. To obtain the sum of the unobscured and obscured SFR of a galaxy, the uncorrected UV SFR is combined with the IR luminosity. The IR luminosity is often only measured at $24\micron$ and the total IR luminosity is estimated using a template spectrum. However, as this method is comparable to fitting two photometric data points by a single galaxy template with only the amount of dust as a free parameter, it will likely result in large uncertainties.

To derive accurate stellar population properties from UV-to-IR SEDs, it is important to explore the full possible range in stellar populations. Fortunately, SPS models that incorporate both the attenuated stellar and dust emissions have recently become available \citep{conroy09,noll09,dacunha08}. These models assume that the energy of the attenuated stellar light is reradiated in the IR. 

In this {\it Letter}, we extend the UV-to-NIR composite SEDs by \citet{kriek11} to MIR wavelengths, and simultaneously fit the stellar and dust emission with the updated Flexible SPS (FSPS) models by \citet{conroy09}. We adopt the following cosmological parameters: $(\Omega_{\rm m}, \Omega_{\Lambda}, h) =$ (0.27, 0.73, 0.7).

\section{Data}

In this work we make use of the composite NUV-to-NIR SEDs by \citet{kriek11}, which were constructed using the photometric catalogs from the NMBS \citep{whitaker11}. The NMBS is a survey in the COSMOS \citep{scoville07} and AEGIS \citep{davis07} fields, which uses five medium-bandwidth NIR filters in the wavelength range $1-1.8 \micron$ designed for NEWFIRM \citep{autry03} on the Mayall 4-m telecope \citep{vandokkum09}. The NIR medium-band photometry has been combined with the publicly available data at NUV-to-NIR wavelength as described in \citet{whitaker11}.

The original composite SEDs were constructed as follows. First, galaxies at $0.5\lesssim z\lesssim 2.0$ with S/N$_{\rm K-band} > 25$ were classified into spectral types based on similarities in their NUV-to-NIR rest-frame SEDs. The number of galaxies in each type varies between 22 and 455. Next, the SEDs of individual galaxies in each type were de-redshifted and scaled to the same reference frame. Finally, the flux was averaged in wavelength bins. See \citet{kriek11} for more details on the procedure. This technique resulted in 32 composite SEDs, which include $\sim3500$ galaxies.

Here, we extend these SEDs to rest-frame MIR wavelengths by adding MIPS $24\micron$ data. Galaxies with active galactic nuclei (AGNs) tend to have a warm dust component at MIR wavelengths \citep{fritz06,feltre12}. Since we are interested in studying dust emission from reprocessed stellar light, we want to avoid contamination by AGNs. Therefore, any galaxies that have detected $L_X \geq 10^{42}$ erg/s in the Chandra COSMOS Survey \citep{elvis09} have been removed. Furthermore, we reject galaxies which are identified to host obscured AGNs based on their IRAC colors, following the criteria by \citet{donley12}. As a result, our sample has been reduced by $\sim$3\%.

\begin{figure}[!ht]
\figurenum{1}
\label{fig:deblending}
\centering
  \begin{tabular}{cccc}  
    \includegraphics[width=.112\textwidth]{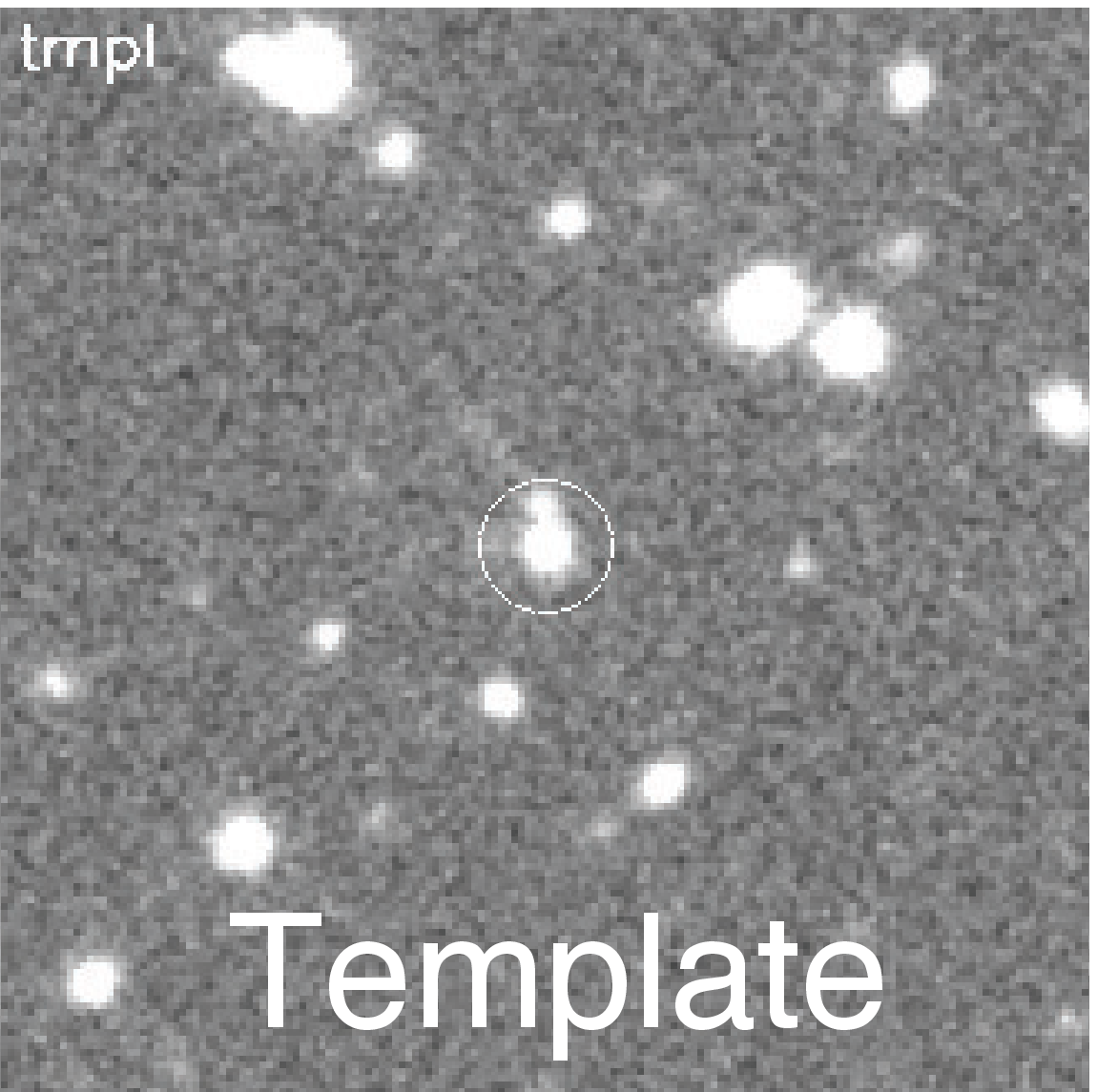} \hspace{-1.2em} &
    \includegraphics[width=.112\textwidth]{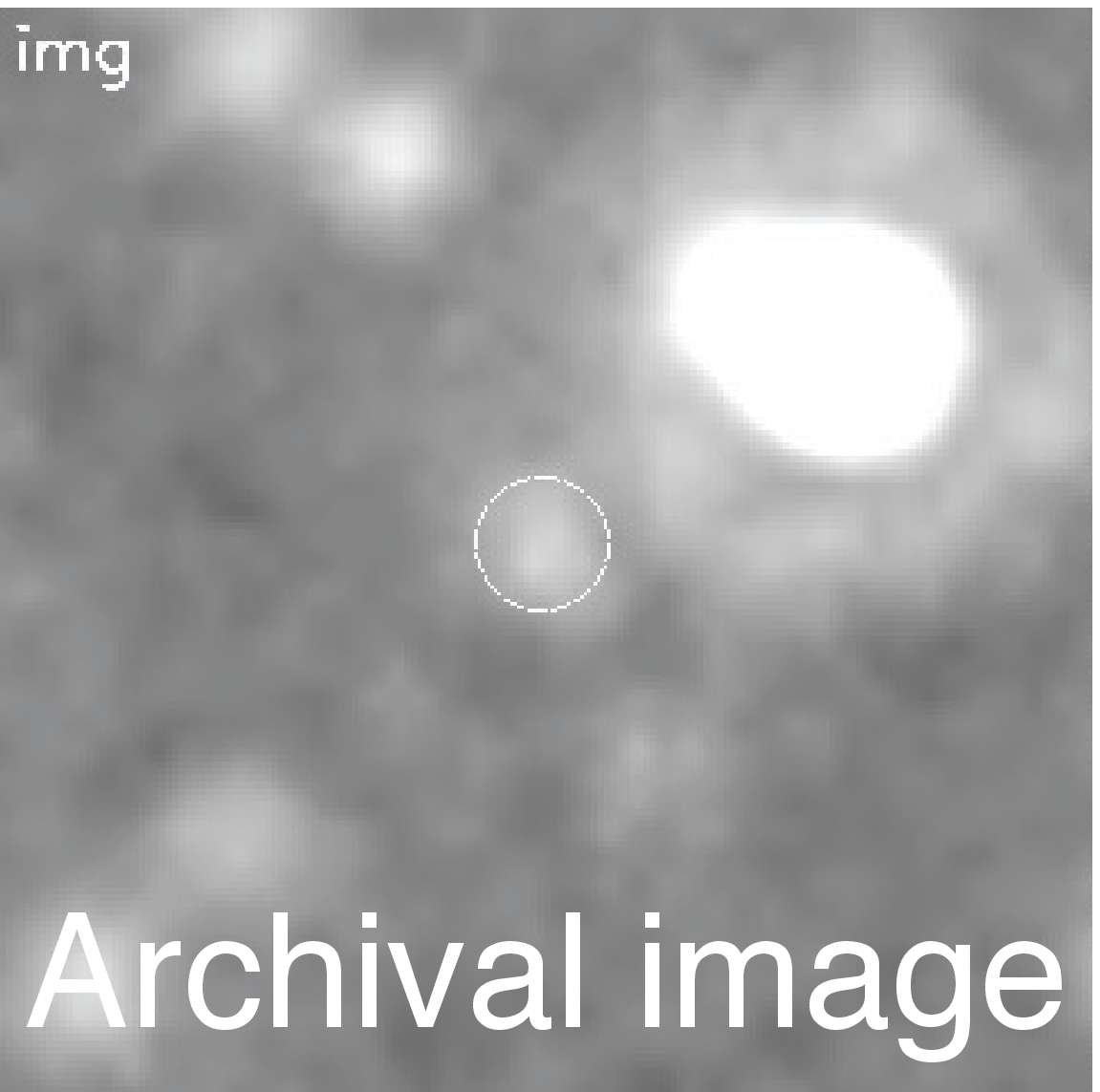} \hspace{-1.2em} &
    \includegraphics[width=.112\textwidth]{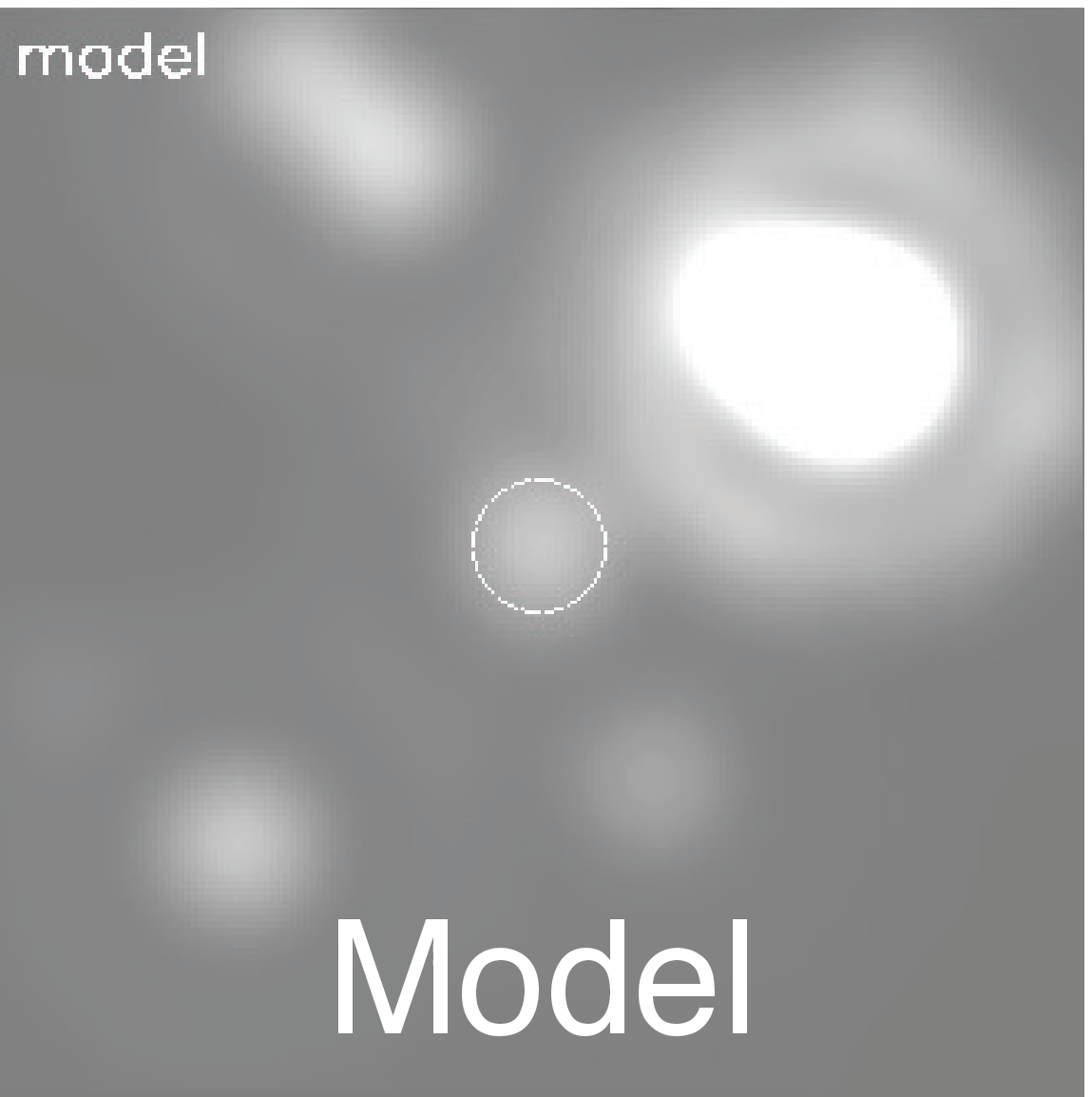} \hspace{-1.2em} &
    \includegraphics[width=.112\textwidth]{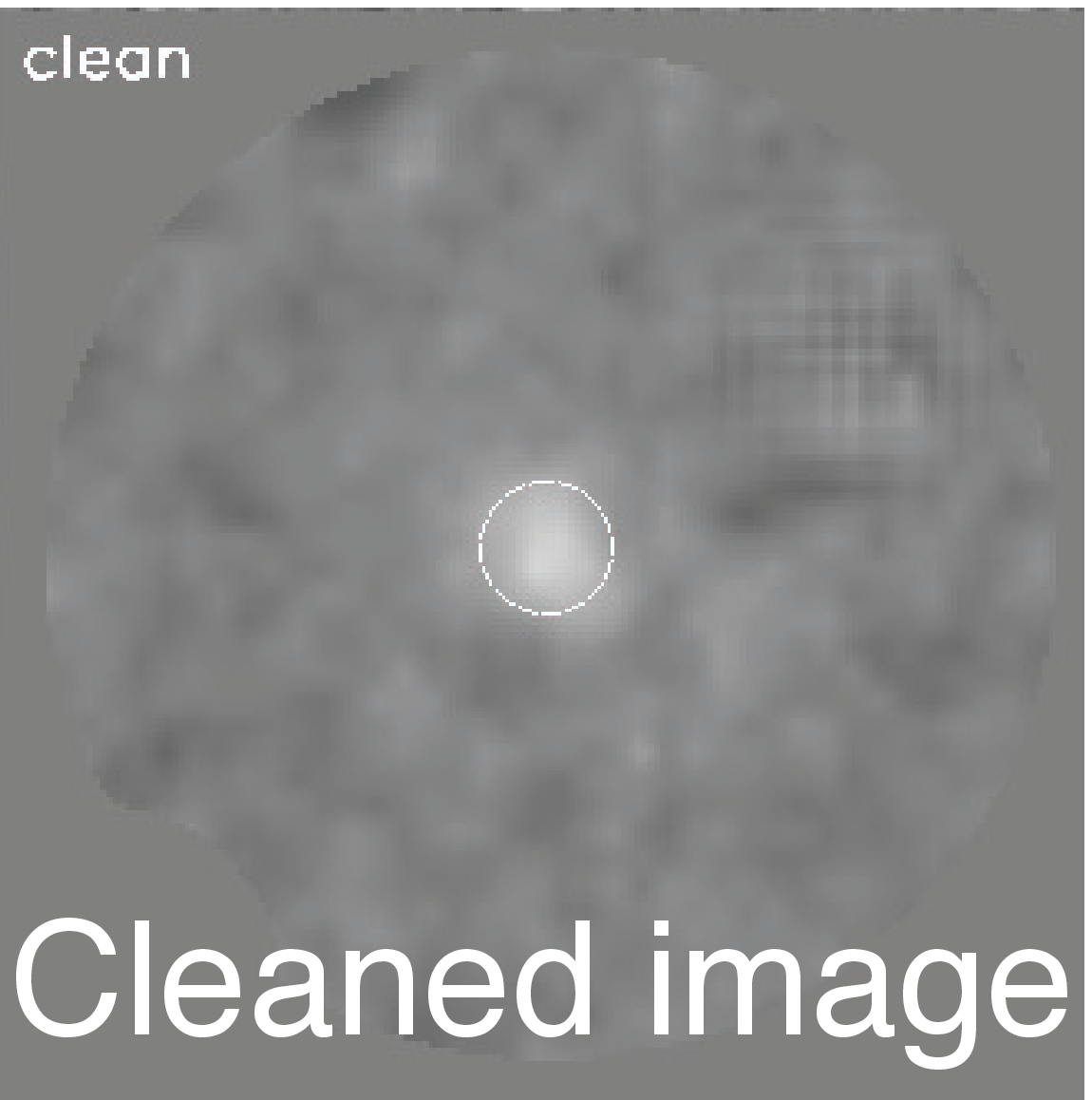} \\ \hspace{-1.2em} \\
    \hline \hspace{-1.2em} \\
    \includegraphics[width=.112\textwidth]{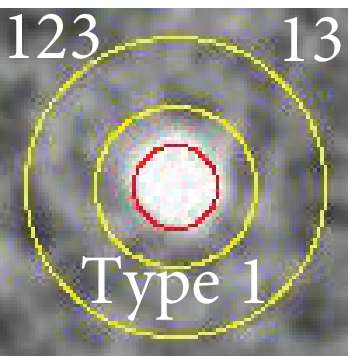} \hspace{-1.2em} &
    \includegraphics[width=.112\textwidth]{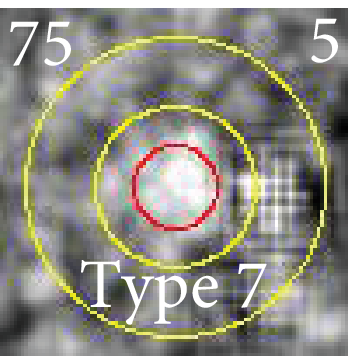} \hspace{-1.2em} &
    \includegraphics[width=.112\textwidth]{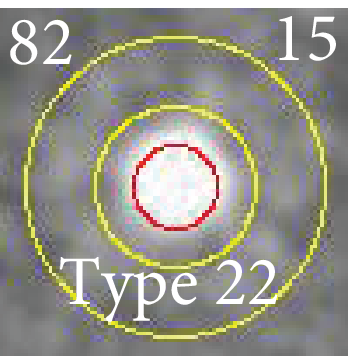} \hspace{-1.2em} &
    \includegraphics[width=.112\textwidth]{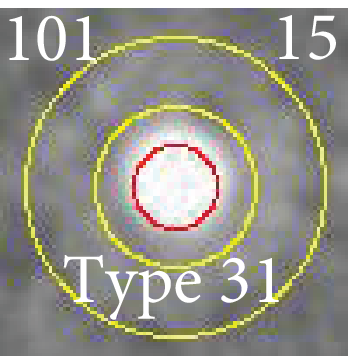} \\ 
  \end{tabular}
\caption{{\it Top:} Source deblending technique. First, the sources are identified using SExtractor from the K-band template image. Next, source models are constructed from the MIPS and K-band PSFs. Finally, all sources except the primary target are removed by subtracting their scaled modeled fluxes. {\it Bottom:} A selection of resulting stacked images ordered by SEDs type. The photometry is measured inside the inner circle, while the background is calculated between the two outer circles. The values in the top left and right corners are the number of galaxies and S/N, respectively.}
\end{figure}

The simplest method to extend the composite SEDs is taking the average of the scaled $24\micron$ fluxes from the NMBS catalogs. However, many sources are undetected, and thus, we stack the images per SED type to obtain deeper photometry. For this purpose, we use the archival mosaic image from the MIPS S-COSMOS Survey \citep{sanders07}. In order to remove contamination by surrounding sources, the MIPS image of each individual galaxy has been cleaned before stacking, using the following steps. First, a model is constructed for all surrounding sources, using the higher resolution K-band image. The K-band image is convolved by a convolution curve derived from the point spread functions (PSFs) of the K-band and MIPS data. Next, we subtract the modeled fluxes of all surrounding sources, to get clean images with a radius of $\sim40\arcsec$ \citep{labbe10}. This technique is illustrated in Figure~\ref{fig:deblending}. The remaining background has been removed by subtracting the average flux within a $7-13\arcsec$ annulus (yellow circles in Figure~\ref{fig:deblending}) from these cleaned images. Then, the cleaned images for each type are stacked into one image, weighted by the scaling factors that are used in the NUV-to-NIR SEDs. We perform a final background subtraction from the stacked image.

The total flux is measured for each stacked image inside a $3.5\arcsec$ aperture (red circles in Figure~\ref{fig:deblending}) and is corrected for missing flux outside the $3.5\arcsec$ aperture using the aperture correction factor from the MIPS instrument handbook. The flux errors are derived using bootstrap resampling of the individual galaxies within the bins.

The stacked MIPS fluxes (spanning $\sim 8 - 15 \micron$ rest-frame) are combined with the NUV-to-NIR data to construct NUV-to-MIR composite SEDs. Composite filter curves are  constructed for the added $24 \micron$ data points, by adding the normalized and de-redshifted $24 \micron$ filter curves for each individual galaxy. Errors on the effective wavelengths are derived by bootstrap resampling the individual galaxies within the bins.

\vfill

\section{SED Fitting}
\label{sec:fitting}

\begin{figure}[!ht]
\figurenum{2}
\label{fig:fitting}
\epsscale{1.21}
\plotone{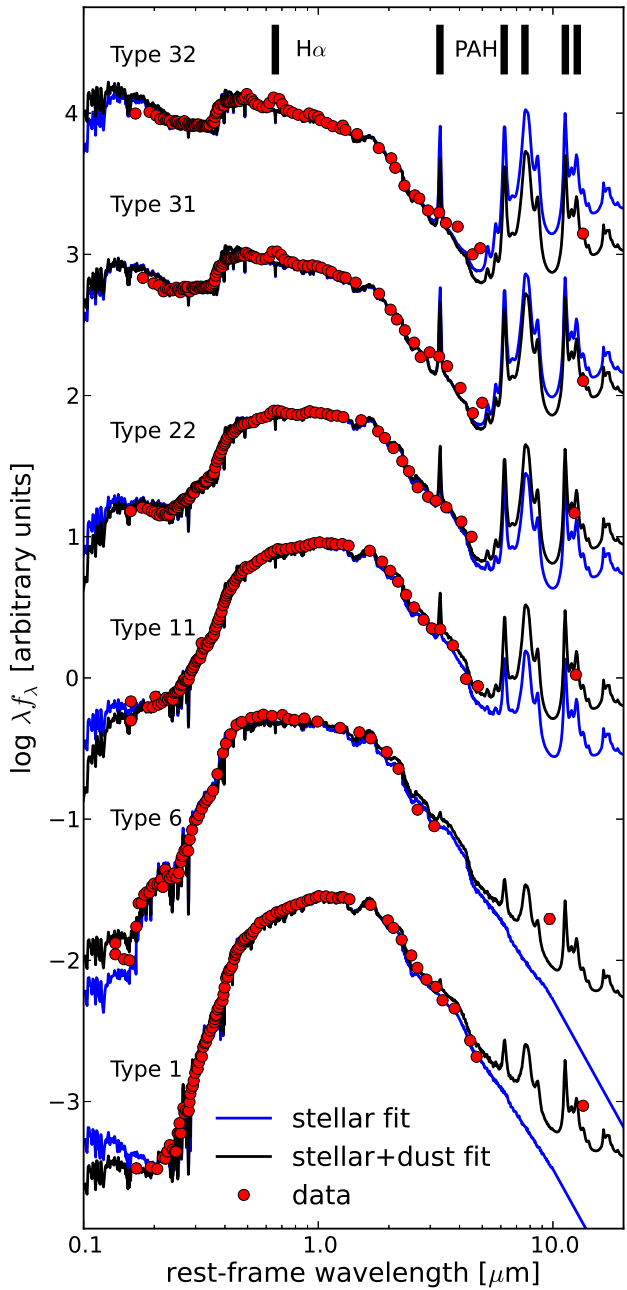}
\caption{A selection of NUV-to-MIR composite SEDs, ordered by decreasing D(4000) strength which is a measure of the age of the stellar population. Red circles represent the stacked data. The best fits to the stellar and stellar+dust emission are represented by the blue and black curves, respectively. The error bars are smaller than the data points. Consistent with previous studies, the expected MIPS fluxes from the `stellar fitting' differ from the observed fluxes. Data points affected by the H$\alpha$ emission line, as clearly seen in star-forming galaxy types, are excluded in the fitting process.}
\end{figure}

\begin{figure}[!ht]
\figurenum{3}
\label{fig:excess_flux}
\epsscale{1.2}
\plotone{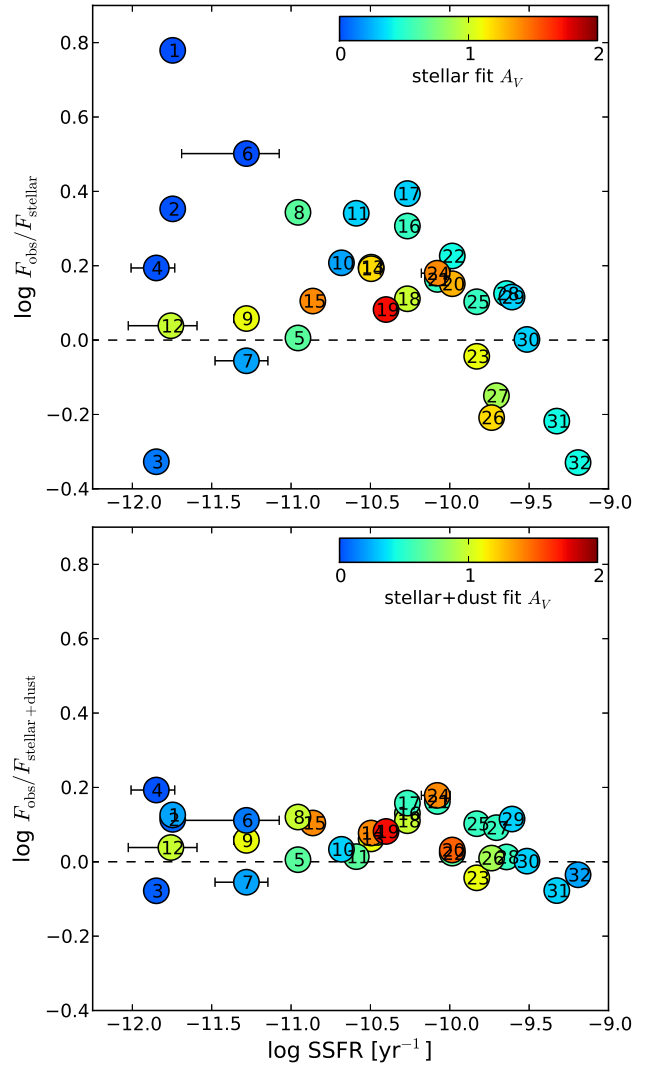}
\caption{Ratio of observed to expected $24 \micron$ flux vs. SSFR for stellar fitting {\it (top)} and stellar+dust fitting {\it (bottom)}. The SSFRs on the x-axis are based on the stellar+dust fitting. Color indicates the dust attenuation in the $V$-band ($A_V$) based on the stellar {\it (top)} and stellar+dust {\it (bottom)} best-fit models. The numbers are the SED types which are ordered by decreasing value of D(4000).}
\end{figure}

In order to derive physical properties, we fit the composite SEDs with the FSPS models \citep{conroy09,conroy10}. We use the BaSeL spectral library \citep{lejeune97,lejeune98,westera02}, Padova isochrones \citep{bertelli94,girardi00,marigo08}, and dust emission models of \citet{draine07}.
Motivated by the work by \citet{kriek13}, which was based on the same composite SEDs, we assume a dust attenuation curve with $R_V = 4.05$ and a UV dust bump which is 20\% of the strength of the Milky Way bump. The three parameters of the \citet{draine07} dust emission model;  $U_{\rm min}$ (specifies the minimum radiation field strength in units of the Milky Way value), $\gamma$ (specifies the relative contribution of dust heated at $U_{\rm min}$ and at $U_{\rm min}\le U\le U_{\rm max}$), and $q$ (the fraction of grain mass in Polycyclic Aromatic Hydrocarbon (PAH) form), are set to their default values, i.e. 0.01, 1.0, and 3.5\%, respectively.

We also assume a delayed-$\tau$ SFH of the form SFR $\propto t \ {\rm exp}(-t/\tau)$ and a \citet{kroupa01} initial mass function (IMF).  The star formation timescale ($\tau$), age, and dust extinction ($A_V$) are left as free parameters, with a minimum log ($\tau$/yr) and log (age/yr) of 7.5. The metallicity is assumed to be Solar.

The fitting is done by minimizing
\begin{equation}
\chi^2 = \sum_i \frac{(F_i - a T_i)^2}{{\delta F_i}^2},
\end{equation}
where
\begin{equation}
a = \frac{\sum F_i T_i / {{\delta F_i}^2}}{\sum T_i^2 / {{\delta F_i}^2}}
\end{equation}
is the scaling factor between the observed flux $F_i$ and the template flux $T_i$ of the model libraries. The template flux is calculated by convolving the flux with the composite filter curves. The flux errors ${\delta F_i}$ are set to be 5\% of the fluxes $F_i$, to avoid that very small flux errors dominate the fit, and to ensure that all data points have equal weight in the $\chi^2$-calculations.

\begin{figure*}[!ht]
\figurenum{4}
\label{fig:sfr}
\epsscale{1.0}
\plotone{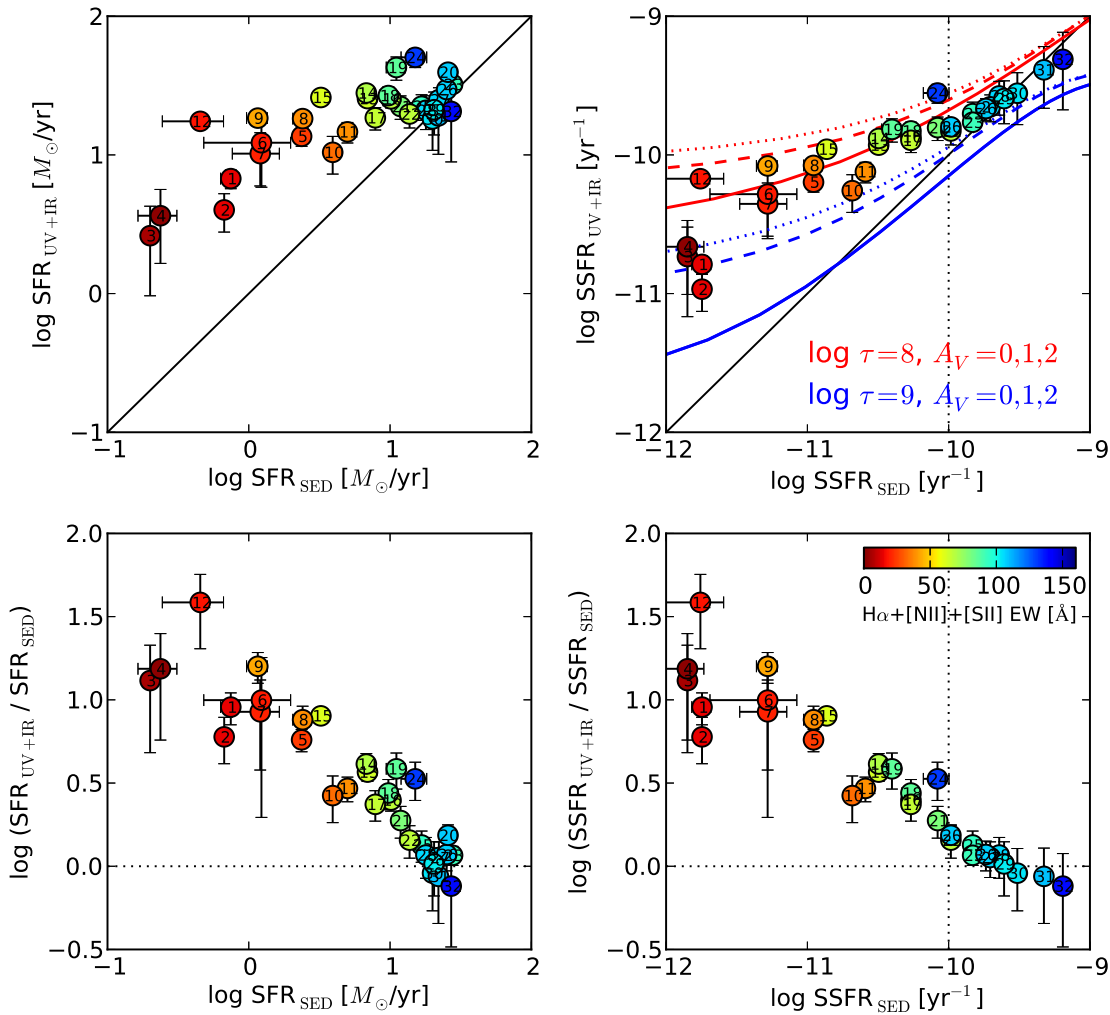}
\caption{{\it Top:} Comparison between (S)SFR based on $L_{\rm UV}$ and $L_{\rm IR}$ vs (S)SFR based on the stellar+dust SED fitting. Solid black lines are one-to-one relationships. {\it Bottom:} Ratio of the two SFRs (SSFR) plotted against SFR (SSFR) based on SED fitting. All plots are color coded with H$\alpha$ equivalent width, and numbered with SED types similar as in Figure~\ref{fig:excess_flux}. Curves in the top right figure are the evolutionary models of the SSFR for a delayed exponential model with $\tau = 8$ and 9 for the red and blue curves, and $A_V = 0, 1$ and 2 for the solid, dashed, and dotted curves, respectively. Only galaxies with log SSFR$_{\rm SED}\gtrsim -10$ lie close to one-to-one relationship.}
\end{figure*}

We calibrate the confidence intervals using Monte Carlo simulations, where the fluxes are perturbed according to a Gaussian distribution, and determine the best-fit parameters. We run 200 simulations and determine the $\chi^2$-level that encloses 68\% of the simulation's best-fits \citep[e.g.,][]{papovich01,kriek09}. Error bars in all figures correspond to these confidence intervals.

In order to get the absolute SFR, we multiply the instantaneous specific SFR, derived from the SED-fitting, by the average stellar mass of the galaxy type. The average mass is derived by assuming the same M/L for all individual galaxies within one type.

The results are shown in Figure~\ref{fig:fitting} for a selection of SED types which range from star-forming, to post-starburst, to quiescent galaxy types. The SEDs are fitted in two ways; by excluding and including the MIPS fluxes. We refer to the former as `stellar fitting' and the latter as `stellar+dust fitting'. For few star-forming and young galaxy types, the NUV region does not have an excellent fit. As shown in \citet{kriek13} better fits can be obtained by allowing both the dust slope and the UV bump strength to vary. However, as this would not significantly change the results of this paper, and would make the fitting impractical, we have decided to fix the dust attenuation law.

We compare the observed $24\micron$ fluxes with the expected fluxes, based on the best-fit stellar and stellar+dust models. The ratios of the observed to expected fluxes are shown in Figure~\ref{fig:excess_flux}. The top panel of Figure~\ref{fig:excess_flux} illustrates that the observed MIPS fluxes become larger than the expected stellar fit model fluxes with decreasing star formation activity. The difference between the observed and modeled fluxes is much smaller ($<0.2$ dex) for the stellar+dust fit (bottom panel of Figure~\ref{fig:excess_flux}). There is no correlation with $A_V$. This result demonstrates the importance of including dust emission while modeling galaxy SEDs, as just modeling the stellar emission may lead to systematic biases in the derived stellar population properties.

\section{Star Formation Rates}
\label{sec:sfr}

With the introduction of MIPS, it has become practice to measure SFRs using both the unobscured light from young stars in the UV and the dust obscured and reprocessed stellar light at IR wavelengths. As MIPS is most sensitive at $24 \micron$, the full IR luminosity is often derived by extrapolating this one data point using a single average galaxy template \citep[e.g.,][]{franx08,wuyts11}. Here, we assess these SFRs using our best-fit models to the dust and stellar emission. 

\begin{figure*}[!ht]
\figurenum{5}
\label{fig:excess}
\epsscale{1.2}
\plotone{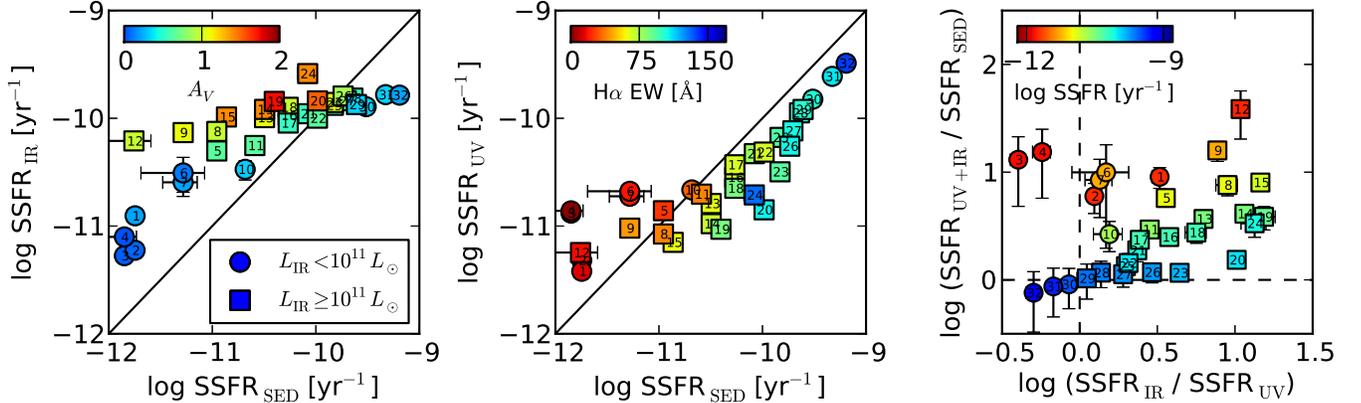}
\caption{SSFR based on IR {\it (left panel)} and UV {\it (central panel)} are plotted versus SSFR based on SED fitting. Both IR and UV-based SSFRs overpredict SED-based SSFR for galaxies with low SSFR. {\it Right panel:} The ratio of UV+IR-based SSFR to SED-based SSFR vs. the ratio of the IR to UV SSFR. Color indicates dust extinction $A_V$ {\it (left panel)}, H$\alpha$ equivalent-width {\it (central panel)}, and log SSFR from SEDs fitting {\it (right panel)}. The data are divided into two categories by \citet{daddi07} based on its $L_{\rm IR}$ {\it (round and square markers)}. The SFR$_{\rm UV+IR}$ overestimates SFR$_{\rm SED}$ by more than $\sim$1\,dex for quiescent galaxies, while for the highest star-forming galaxies the two SFRs are broadly consistent.}
\end{figure*}

We use the monochromatic conversion template by \citet{wuyts08} to infer $L_{\rm IR}$ from $24\,\micron$ flux. This template is a luminosity independent template, derived by taking the log average of the exponents of the interstellar radiation field strength in the \citet{dale02} templates. \citet{wuyts11} found that this template is well matched with the SFR of star-forming galaxies based on UV+PACS data in the range of $0 < z < 3$. However, the accuracy of this template for quiescent and transition galaxies has not been assessed.

Next, the total SFR based on both UV+IR can be calculated by \citep{bell05,kennicutt98}:
\begin{equation}
\label{eq:sfr}
{\rm SFR}_{\rm UV+IR} \ [M_\odot{\rm /yr}] = 9.8 \times 10^{-11} \ (L_{\rm IR} + 2.2 \ L_{\rm UV}),
\end{equation}
assuming a \citet{kroupa01} IMF and luminosity in $L_{\odot}$. Here, $L_{\rm UV}$ is defined as $1.5 \ \nu L_{\nu}$ at 2800 \AA, which is a rough estimate of the total integrated $1216 - 3000$ \AA\ UV luminosity, and the factor of 2.2 accounts for the unobscured light of young stars that is emitted outside the $1216 - 3000$\,\AA\ band \citep{bell05}. Note that $L_{\rm UV}$ is derived using the flux interpolation at 2800 \AA. Thus, this method basically fits two data points of the full SED with one star-forming galaxy template, with only the amount of obscuration as a free parameter. While this method has been calibrated using active star-forming galaxies, it would not be surprising if it breaks down for galaxies that have different stellar populations.

The comparison between (S)SFR$_{\rm SED}$ and (S)SFR$_{\rm UV+IR}$ are shown in Figure~\ref{fig:sfr}. Generally, for galaxies with older stellar populations (i.e., lower SSFR), SSFR$_{\rm UV+IR}$ is higher than SSFR$_{\rm SED}$, while younger and higher SSFR galaxies lie closer to one-to-one relation. In order to assess whether this discrepancy may be due to the fact that we use 2800\,\AA\ instead of 1600\,\AA, we calculate the expected SFRs using 1600 \AA \ best-fit model fluxes and 24$\micron$ observed flux, but still find similar results. \citet{wuyts11} found that ${\rm SFR}_{\rm UV+IR}$ overestimates SFR$_{\rm SED}$, in particular for high SFRs, when short star formation timescales were allowed. Our results do not change significantly when we restrict the star formation timescale to log $\tau > 8.5$ or when we assume an exponentially declining SFH. Thus, we argue that SSFR$_{\rm UV+IR}$ overestimates  SSFR$_{\rm SED}$ for galaxies with log\,SSFR $\lesssim -10$, and the discrepancy becomes larger with decreasing SSFR.

\section{Discussion}

We find that SFR$_{\rm UV+IR}$ overestimates SFR$_{\rm SED}$ by more than $\sim 1$\,dex for quiescent galaxies, while for the most active star-forming galaxies in our sample the two SFRs are broadly consistent (Figure~\ref{fig:sfr}). Our results are consistent with recent findings by \citet{mattia13} and \citet{salim09}. In order to investigate the cause of the difference between SSFR$_{\rm SED}$ and SSFR$_{\rm UV+IR}$, we dissect the SSFR$_{\rm UV+IR}$ in SSFR$_{\rm UV}$ and SSFR$_{\rm IR}$. We also plot the ratio between the latter two and compare it with the ratio between SSFR$_{\rm UV+IR}$ and SSFR$_{\rm SED}$ (Figure~\ref{fig:excess}). No dust correction was applied to derive SSFR$_{\rm UV}$.

For the majority of types, the SSFR excess is dominated by the MIR flux, while for a few, it is dominated by the UV flux. This SSFR excess can be caused by the contribution of old \citep[e.g.,][]{mattia13} and/or intermediate-age stars \citep[e.g.,][]{salim09,kelson10} to the MIR and UV light, which explains the strong correlation with SSFR. This finding is not surprising, as we only use a single star-forming galaxy template (by \citet{wuyts11}) when estimating SFRs from UV+IR. Therefore, only young and star-forming galaxies with SSFR $\gtrsim 10^{-10}$  yr$^{-1}$ lie close to one-to-one relation (see also \citet{arnouts13}).  In this case, $L_{\rm UV+IR}$ is a robust estimator of the SFR. However, there might be an upper limit where the agreement between the two methods breaks down again, as \citet{wuyts11} found that SFR$_{\rm UV+IR}$ overpredicts  SFR$_{\rm SED}$ at high redshift ($z \gtrsim 2.5$) and at the high-SFR-end ($\gtrsim 100\,M_{\odot}$/yr).

Compton-thick AGNs with $L_X \gtrsim 10^{43}$~erg/s could also explain the discrepancy between SSFR$_{\rm SED}$ and SSFR$_{\rm UV+IR}$, due to their MIR excess \citep{daddi07b}. However, we removed AGNs identified by their strong X-ray flux or by an IRAC upturn \citep{donley12}. Nonetheless, we cannot rule out contributions from low luminosity AGNs, and X-ray stacks of the same composite SED sample indeed indicate low levels of black hole accretion \citep{jones13}. \citet{daddi07} also reported that MIR excess galaxies have $L_{\rm IR} \gtrsim 10^{11} L_{\odot}$. We check this possibility, but we do not find such trend in our data (Fig.~\ref{fig:excess}).

\citet{nordon10} proposed that the discrepancy between SFR$_{\rm SED}$ and SFR$_{\rm UV+IR}$ is due to excess in PAH emission, rather than obscured AGNs. We conduct a test for this hypothesis by varying $\gamma$, $U_{\rm min}$, and $q$. We find that the SFR$_{\rm SED}$ is not sensitive to these variations, but SFR$_{\rm UV+IR}$ can be affected due to the used monochromatic conversion template. Therefore, higher observed-frame $24\micron$ fluxes due to variations in PAH emission can lead to discrepancies between SFR$_{\rm SED}$ and SFR$_{\rm UV+IR}$.

Lastly, we mention that circumstellar dust around AGB star is not yet included in the FSPS models, but is subject of ongoing work. In this context, it is interesting to note that the post-starburst galaxy types (6 \& 7), for which we expect the highest contribution from AGB stars, have an MIR excess, though it is not larger than for the other quiescent galaxy types.

\section{Summary}

In this letter, we use NUV-to-MIR composite SEDs to simultaneously model the stellar and dust emission in distant galaxies. NUV-to-NIR composite SEDs had previously been constructed from the NMBS photometry of $\sim3500$ galaxies at $0.5\lesssim z\lesssim 2$, by matching galaxies with similar SED shapes. In this work, we extend the SEDs with a stacked MIPS\,24\,$\micron$ datapoint, resulting in multi-wavelength SEDs spanning from $\sim0.2$ to $15\micron$ in rest-frame wavelength.

Stellar population properties are derived by fitting the composite SEDs with the FSPS models, which include both stellar and dust emission. Consistent with previous studies, we find that the predicted MIPS flux, based on fitting just the stellar emission, is inconsistent with the observed MIPS flux for most galaxy SED types.

We use the best-fit SFRs from the full stellar and dust fitting to assess SFRs determined from the UV and IR luminosities, currently the most popular method to determine SFRs. We find that (S)${\rm SFR}_{\rm UV+IR}$ overpredicts (S)${\rm SFR}_{\rm SED}$ for galaxies with log ${\rm SSFR}\lesssim-10$, and the discrepancy becomes increasingly larger for lower SSFR. The discrepancy is due to both UV and MIR luminosities, though the MIR is the dominant contributor for most SED types. Contributions from obscured and unobscured old and/or intermediate-age stellar populations to the MIR and UV luminosities are the likely explanation for the overestimated SFR$_{\rm UV+IR}$.

Based on our results, we conclude that SFRs should be determined from modeling stellar and dust emission simultaneously, instead of just measuring the UV and MIR luminosities. An important implication of our work is that quiescent galaxies have even lower SFRs than what was previously found, based on UV and IR luminosities. However, young star-forming galaxies with SSFR $\gtrsim 10^{-10}$ ${\rm yr}^{-1}$ lie close to one-to-one relation, and thus $L_{\rm UV+IR}$ is a robust SFR estimator.

The composite SEDs currently only extend to MIR wavelengths, and thus the SFRs derived from the modeled dust and stellar emission may still suffer from systematics. In future studies we will extend the SEDs to FIR wavelengths, to measure the full bolometric luminosity and more accurately measure the total SFR.

\acknowledgments

We thank the NMBS and COSMOS collaborations for making their catalogs publicly available, and Marijn Franx and Edward Taylor for useful discussions.

\end{document}